# Naming the SMD Bridge Program – A White Paper

Carl A. Moore Jr[1]., Jesús Pando[2]

[1] Department of Mechanical Engineering
Florida A&M University
[2] Department of Physics and Astrophysics
DePaul University

ABSTRACT

NASA's Science Mission Directorate (SMD) has initiated a program to enhance the participation of historically underrepresented institutions and communities in NASA's mission. Currently known as the NASA SMD Bridge Program, its goal is to establish enduring partnerships among these institutions, research-intensive universities, and NASA centers. There are concerns about using "Bridge" in the program's name, with stakeholders suggesting that it might stigmatize students and mislead applicants about its focus. In this white paper, we address these concerns and conclude that a name change that better reflects the mission of this new SMD effort is necessary to address these concerns.

I. Introduction

Traditionally, an educational Bridge is a program that assists students in transitioning to a higher academic level. Different Bridge programs operate with distinct goals and initiatives. However, most aim to provide opportunities such as fellowships, research experience, common applications, and funding to marginalized communities. Some Bridge programs offer supplemental educational and emotional support to students from low-income families to help them successfully navigate the college experience[1]. Another type of Bridge program is a summer experience designed to ease the transition to college and support postsecondary success by providing students with the academic skills and social resources needed to thrive in a college environment[2].

NASA's Science Mission Directorate has launched a new initiative aimed at improving diversity, equity, inclusion, and accessibility (DEIA) within the NASA workforce and the broader U.S. science and engineering communities. The initiative, currently called the NASA SMD Bridge Program, aims to foster sustainable partnerships between institutions that have historically been under-resourced by NASA, such as Minority-Serving Institutions (MSIs) including Historically Black Colleges and Universities (HBCUs), Tribal Colleges and Universities (TCUs), Primarily Undergraduate Institutions (PUIs), Primarily Black Institutions (PBIs), Hispanic Serving Institutions (HSIs), Community Colleges, and highly research-intensive universities and NASA Centers or Facilities[3].

The SMD established a Bridge Program Workshop Organizing Committee (BPWOC) to gather ideas and suggestions from the community stakeholders on what the SMD Bridge Program Announcement of Opportunity (AO) should include. The BPWOC conducted a week-long workshop that raised two concerns about using the word "Bridge." These concerns and many other guiding themes and perspectives that shape the program are documented in the report, which is publicly available on the BPWOC report [website] [4]. Firstly, some participants highlighted that traditional bridge programs have negative connotations as they focus on students' weaknesses rather than strengths. Second, many noted that the goals of traditional bridge programs do not capture the spirit of the SMD program. To address these concerns, the BPWOC formed a subcommittee to investigate the impact of "Bridge" in the program's title.

**Concern I—Stigmatization of Students**

The subcommittee interviewed Arlene Modeste Knowles, the TEAM-UP Diversity Task Force Project Manager at the American Institute of Physics, who is knowledgeable about bridge programs in physics and astronomy. The conversation confirmed that bridge programs have successfully graduated students of color. The published literature on Summer Bridge and Bridge to the PhD programs supports this conclusion [5,6]. However, some program graduates reported feeling like second-class citizens because their program separated them from or placed them on a different path than the traditionally accepted graduate students. Additionally, to ensure students' success, some bridge programs have admitted students who could have been accepted into the graduate programs outright. For these students, a remedial course or two may have been sufficient to catch them up, and they would not have been delayed in their matriculation or placed in a subgroup outside of the normal first-year graduate cohort. Therefore, some believe that administrators may place some students of color in bridge programs even though a proper evaluation may have indicated their ability to enter as traditional graduate students.

The subcommittee has proposed renaming the NASA SMD Bridge Program for three reasons. First, the program has a unique focus that sets it apart from traditional bridge programs. Second, some bridge program graduates feel they were treated as second-class students. Third, the term "bridge program" has gained a negative connotation among some communities it intends to serve.

**Concern II—Scope of the Program**

To better understand the landscape of traditional Bridge programs and investigate the difference between their goals and the SMD program, the BPWOC surveyed the following programs:

- Bridge to the Ph.D. Program in STEM - Columbia University[7]
- Cal-Bridge - Cal State and UC Schools[8]
- Fisk-Vanderbilt Master's-to-PhD Bridge Program - Fisk and Vanderbilt University[9]
- APS Bridge Program - American Physical Society[10]
- AGU Bridge Program - American Geophysical Union[11]
- Bridge to the Doctorate - University of Virginia[12]
- ACS Bridge Project - American Chemical Society[13]

- The BRIDGE program - University of Pittsburgh[14]
- Penn's Bridge to Ph.D. - University of Pennsylvania[15]

These programs have varying application requirements, funding support, academic components, and levels of success. Still, their primary objective is to assist underrepresented students in obtaining Ph.D. degrees in STEM fields.

The purpose of the SMD program is not solely to increase opportunities for traditionally marginalized students. The workshop resulted in two core tenets. Firstly, the ideal NASA Bridge Program would center the needs of students and include historically and systematically marginalized faculty and institutions. Secondly, the ideal NASA Bridge Program would lead a paradigm shift by assuming primary responsibility for building impactful relationships and partnerships with marginalized and underserved communities to diversify its workforce and the STEM community [4].

The workshop working groups arrived at several key findings aligned with the two core tenets. Some of these include [4].
- Enable/Support Collaboration Building and Cohort Formation
- Appoint and Support NASA Facilitator(s)/'Point People' to be Responsible for Making and Sustaining Connections with Under-served Institutions
- Provide Realistic Financial Support and Keep the Majority of Funding at/in Underserved Institutions. Create Support and Structures for Developing, Maintaining, and Evaluating Effective Mentoring at Participating Sites
- Provide Support and Training for Early Career Professionals at Underserved Institutions, as well as Opportunities for Research Growth

The findings suggest that the SMD program has a different focus than traditional bridge programs. It aims to address issues of systemic underrepresentation by creating opportunities for institutions, faculties, and students who have not traditionally engaged with NASA, with under-resourced institutions being the primary partner. While the program aims to enhance the success of marginalized students, it also aims to enable institutions to participate successfully in NASA's enterprise. This program's focus is broader in scope and scale than traditional bridge programs.

**Conclusion**

The subcommittee has concluded that the SMD Bridge Program has a wider scope than a typical bridge program. While a typical bridge program aims to help students transition to higher academic levels, the SMD Bridge Program aims to create sustainable partnerships between NASA SMD and historically underrepresented universities, schools, colleges, faculty, and students. It is important to market this program clearly to ensure that these stakeholders know the opportunities available. As part of this marketing effort, the subcommittee recommends that the program name not include "Bridge" as it may carry negative connotations and discourage some from exploring the program's possibilities. Instead, the program's name should more accurately reflect the full range of access points the program will offer.

**References**

[1] Grace-Odeleye, B., & Santiago, J. (2019). A Review of Some Diverse Models of Summer Bridge Programs for First-Generation and At-Risk College Students. *Administrative Issues Journal: Connecting Education, Practice, and Research*, *9*(1), 35-47.

[2] Kallison Jr, J. M., & Stader, D. L. (2012). Effectiveness of summer bridge programs in enhancing college readiness. *Community College Journal of Research and Practice*, *36*(5), 340-357.

[3] Program details - https://science.nasa.gov/researchers/smd-bridge-program/

[4] https://smd-cms.nasa.gov/wp-content/uploads/2023/11/reportfromthebridgeprogramworkshoporganizingcommittee.pdf?utm_source=TWITTER&utm_medium=NASAScienceAA&utm_campaign=NASASocial&linkId=248811594

[5] Bradford, B. C., Beier, M. E., & Oswald, F. L. (2021). A meta-analysis of university STEM summer bridge program effectiveness. *CBE—Life Sciences Education*, *20*(2), ar21.

[6] Rudolph, A. L., Holley-Bockelmann, K., & Posselt, J. (2019). PhD bridge programmes as engines for access, diversity and inclusion. *Nature Astronomy*, *3*(12), 1080-1085.

[7] https://bridgetophd.facultydiversity.columbia.edu

[8] https://www.cpp.edu/calbridge/index.shtml

[9] https://www.fisk-vanderbilt-bridge.org

[10] https://www.aps.org/programs/minorities/bridge/index.cfm

[11] https://www.agu.org/learn-and-develop/learn/travel-research-grants/agu-bridge-program

[12] https://graduate.as.virginia.edu/bridge-doctorate

[13] https://www.acs.org/education/students/graduate/bridge-project.html

[14] https://www.engineering.pitt.edu/programs/bridge-program/bridge-program/

[15] https://gsc.upenn.edu/bridge-phd-program-provides-way-forward-greater-access-stem-fields